\documentclass[a4,12pt]{article}

\usepackage[latin1]{inputenc}

\newcommand{\jmu}{\eta}

\newcommand{\varphihere}{\varphi}

\newcommand{\MUone}{M/\mathrm{U(1)}}%
\newcommand{\Uone}{\mathrm{U(1)}}%

\usepackage{color}
\usepackage{graphicx}  % standard LaTeX graphics tool

\usepackage{psfrag}

\usepackage{tocbibind} %adds contents and bibliography to the table of contents

\usepackage{epsf}
\usepackage{epic}
\usepackage{eepic}

\usepackage{epsfig}
\usepackage{wrapfig}
\usepackage{amsmath,amssymb,amsfonts}
\usepackage[mathscr]{eucal}
\usepackage{citesort}
\usepackage[active]{srcltx}
\usepackage{url}
\usepackage{fancybox}
\usepackage{ntheorem}

\usepackage{mhequ}

\newcommand {\twist} {\lambda}

\newcommand{\myomega}{v}

\newcommand{\beq}{\begin{equation}}

\newcommand{\FS}       %{F_1} %
                  {F}
\newcommand{\HS} %{F_2}
       {H_{\mbox{\scriptsize volume}}}

{\ptc{this should be removed in the oberwolfach version}}%

\newcommand{\mcA}{{\mycal A}}%
\newcommand{\eeal}[1]{\label{#1}\end{eqnarray}}
\newcommand{\bed}{\begin{deqarr}}
\newcommand{\eed}{\end{deqarr}}
\newcommand{\bedl}[1]{\begin{deqarr}\label{#1}}
\newcommand{\eedl}[2]{\arrlabel{#1}\label{#2}\end{deqarr}}

\newcommand{\bel}[1]{\begin{equation}\label{#1}}
\newcommand{\bea}{\begin{eqnarray}}
\newcommand{\bean}{\begin{eqnarray}\nonumber}
\newcommand{\beal}[1]{\begin{eqnarray}\label{#1}}
\newcommand{\eea}{\end{eqnarray}}

\newcommand{\be}{\begin{equation}}
\newcommand{\eeq}{\end{equation}}
\newcommand{\ee}{\end{equation}}
\newcommand{\beqa}{\begin{eqnarray}}
\newcommand{\eeqa}{\end{eqnarray}}
\newcommand{\beqan}{\begin{eqnarray*}}
\newcommand{\eeqan}{\end{eqnarray*}}
\newcommand{\ba}{\begin{array}}
\newcommand{\ea}{\end{array}}

\newtheorem{Theorem} {\sc  Theorem\rm} [section]

\theorembodyfont{\upshape}
\newtheorem{Remark}[Theorem]{\sc Remark\rm}

\theoremsymbol{\ensuremath{\diamondsuit}}
\newcommand{\fcoco}{\small}
\theorembodyfont{\fcoco} \theoremseparator{} \theoremindent0.5cm

\theoremstyle{nonumberplain} \theorembodyfont{\fcoco}
\theoremseparator{} \theoremindent0.5cm

\DeclareFontFamily{OT1}{rsfs}{}
\DeclareFontShape{OT1}{rsfs}{m}{n}{ <-7> rsfs5 <7-10> rsfs7 <10->
rsfs10}{} \DeclareMathAlphabet{\mycal}{OT1}{rsfs}{m}{n}

{\catcode `\@=11 \global\let\AddToReset=\@addtoreset}
\AddToReset{equation}{section}
\renewcommand{\theequation}{\thesection.\arabic{equation}}

\newcounter{mnotecount}[section]

\renewcommand{\themnotecount}{\thesection.\arabic{mnotecount}}

\newcommand{\mnote}[1]%{}%
{\protect{\stepcounter{mnotecount}}$^{\mbox{\footnotesize
$%\!\!\!\!\!\!\,
\bullet$\themnotecount}}$ \marginpar{%\color{red}%
\raggedright\tiny\em
$\!\!\!\!\!\!\,\bullet$\themnotecount: #1} }

\newcommand{\warn}[1]%{}%{}
{\protect{\stepcounter{mnotecount}}$^{\mbox{\footnotesize
$%\!\!\!\!\!\!\,
\bullet$\themnotecount}}$ \marginpar{%\color{red}%
\raggedright\tiny\em $\!\!\!\!\!\!\,\bullet$\themnotecount: {\bf
Warning:} #1} }

\newcommand{\R}{\mathbb R}

\newcommand{\eq}[1]{(\ref{#1})}

\newcommand{\Mext}{M_\ext}
\newcommand{\ext}{\mathrm{ext}}

\newcommand{\ptc}[1]{\mnote{#1}}

\newcommand{\mcL}{{\mycal L}}

\newcommand{\beqar}{\begin{deqarr}}
\newcommand{\eeqar}{\end{deqarr}}

\newcommand{\beaa}{\begin{eqnarray*}}
\newcommand{\eeaa}{\end{eqnarray*}}

\newcommand{\tr}{\mbox{tr}}

\title{Mass, angular-momentum, and charge inequalities for  axisymmetric initial
data}

\author{Piotr T. Chru\'sciel\\
LMPT, Fédération Denis Poisson,  Tours
\\
Mathematical Institute and Hertford College, Oxford
\\
\\
Jo\~ao Lopes Costa
\\
Lisbon University Institute -- ISCTE
\\
Mathematical Institute and Magdalen College, Oxford
 }

\begin{document}
\maketitle

\begin{abstract}
We present the key elements of the proof of an upper bound for
angular-momentum and charge in terms of the mass for
electro-vacuum asymptotically flat axisymmetric initial data
sets with simply connected orbit space.
\end{abstract}

%\input{kerr}
%
%\end{document}
\section{Introduction}
\label{Sintro}
 In important recent work, Dain~\cite{Dain:2006} has
proved an upper bound for angular-momentum in terms of the mass for
a class of maximal, vacuum, axisymmetric initial data
sets.\footnote{The analysis of~\cite{Dain:2006} has been extended
in~\cite{CLW} to include all axisymmetric vacuum initial data, with
simply connected orbit space, and manifolds which are asymptotically
flat in the standard sense, allowing moreover several asymptotic
ends.}
The aim of this work is to generalize Dain's inequality to include
electric and magnetic charges. In fact, the heuristic argument
behind Dain's original inequality generalizes to electro-vacuum as
follows: the ``standard picture of gravitational
collapse"~\cite{Dain:2007} is that the formation of event horizons
should occur {\em generically} in large families of space-times,
with the exterior region approaching a Kerr-Newman metric
asymptotically with time. Now, mass and Maxwell charges are
conserved quantities, and the same is true for angular momentum if
one further assumes axisymmetry, so the inequality~\eqref{21VI.1}
follows for all initial data for such a collapse.  Besides its
intrinsic interest, our inequality provides some support for this
``standard picture", and in particular for {\em weak cosmic
censorship}.

More precisely, consider a three dimensional electro-vacuum smooth
initial data set $(M,g,K, E, B)$, where $M$ is the union of a
compact set and of two asymptotically flat regions  $M_1$ and $M_2$.
Here $g$ is a Riemannian metric on $M$, $K$ is the extrinsic
curvature tensor, $E$ is the electric field and $B$ the magnetic
one, both divergence-free in electro-vacuum.  We suppose that the
initial data set is \emph{axisymmetric}, by which we mean that it is
invariant under an action of $\Uone$, and maximal: $\tr_g K =0$. It
is further assumed that $\MUone$ is simply connected, so that the
results of~\cite{ChUone} can be used. The notion of asymptotic
flatness is made precise in \eq{falloff1} and \eq{KFfalloff1}, where
moreover $k\ge 6$ needs to be assumed when invoking~\cite{ChUone}.
We then have the following:

\begin{Theorem}
 \label{TM1}
Under the conditions just described, let $m$, $\vec J$, $Q_E$
and $Q_B$ denote respectively the ADM mass, the ADM angular
momentum, the total electric charge and the total magnetic
charge of $M_1$:
$$Q_E=-\frac{1}{4\pi}\int_{S_{\infty}}*F\;,
 \quad
 Q_B=\frac{1}{4\pi}\int_{S_{\infty}}F\;.
$$
Then
\bel{21VI.1}
 m\ge \sqrt{\frac{|\vec J|^2}{m^2}+Q_E^2+Q_B^2} \;.
\ee
\end{Theorem}

A slightly more general version of Theorem~\ref{TM1} can be
found in Theorem~\ref{TM2} below.

The reader should note an inequality relating area, angular
momentum, and charge, proved for stationary Einstein-Maxwell
black holes in~\cite{HennigAnsorgCederbaumEM}, as well as  the
discussion of the Penrose inequality in electrovacuum
of~\cite{WeinsteinYamada}.

\begin{Remark}
We expect the equality to be attained only for the magnetically and
electrically charged extreme Kerr-Newman space-times, which are
unlikely to satisfy the hypotheses of Theorem~\ref{TM1}.
\end{Remark}
\begin{Remark}
\label{J=0} If $M$ contains only one asymptotic flat end and
$\partial M=\emptyset$ we have $\vec J=Q_E=Q_B=0$ (see
\eq{Jvanish} below), whence our interest in initial data sets
containing two ends. One expects the result to generalize to
several ends along the lines of~\cite{CLW}, but a proof of this
lies beyond the scope of this work.
\end{Remark}

\begin{Remark}
The proof applies to Einstein--Abelian Yang-Mills fields
configurations, giving in this case
\bel{21VI.2}
 m \ge \sqrt{\frac{|\vec J|^2}{m^2}+(\sum_i Q_{E_i})^2+(\sum_i Q_{B_i})^2}
 \;,
\ee
where the $Q_{E_i}$'s and the $Q_{B_i}$'s are the electric and
magnetic charges associated with the $i$'th Maxwell field.
\end{Remark}

\section{Angular momentum and charge inequalities}
 \label{SAmci}

Recall that an \emph{asymptotically  flat end} is a region
$\Mext\subset M$ diffeomorphic to $\R^3\setminus B(R)$, where
$B(R)$ is a coordinate ball of radius R, such that in local
coordinates on $\Mext$ obtained from $\R^3\setminus B(R)$ we
have, for some $k\geq1$,%
\footnote{We write $f=o_k(r^{-\alpha})$ if the limits $
  \lim_{r\to\infty}r^{\alpha+\ell}\partial_{k_1}\ldots\partial_{k_\ell}
f$ vanish for all $ 0\le \ell \le k $, and $f=O_k(r^{-\alpha})$
if there exists a constant $C$ such that $
  |r^{\alpha+\ell}\partial_{k_1}\ldots\partial_{k_\ell}
f|\le C$  for all $ 0\le \ell \le k $.}
\bel{falloff1} g_{ij}=\delta_{ij}+o_k(r^{-1/2})
 \;, \ \partial_k g_{ij} \in L^2(\Mext)\;, \quad
 K_{ij}=O_{k-1}(r^{-\lambda}) \;,\  \lambda > \frac 52\;.
\ee
(The asymptotic conditions on $g$ arise from the requirement of
well defined ADM mass, with the integrability condition
satisfied if, e.g., $\partial_k g_{ij}=O(r^{-\alpha-1})$, for
some $\alpha>1/2$. The restriction on the decay rate of $K$
above arose already in the vacuum case, and can be traced back
to the unnumbered equation after (2.37) in~\cite{CLW}.)

The electric and magnetic fields $E$ and $B$ are defined to be
the orthogonal projections to $TM$ of their space-time
analogues
\bel{electromagnetic fields}
E^{\mu}=F^{\mu}{}_{\nu}n^{\nu}\;,\quad
B^{\mu}=*F^{\mu}{}_{\nu}n^{\nu}\;, \ee
where $F$ is the Maxwell two-form, and where $n$ is a unit
normal to $M$, when embedded in a space-time. We assume that in
the manifestly asymptotically flat coordinates we have
\bel{KFfalloff1}
 E^i=O_{k-1}(r^{-\gamma-1}) \;, \quad
 B^i=O_{k-1}(r^{-\gamma-1})\;\;,\; \gamma>3/4
% \;,
% \quad
% |E|^2+|B|^2 \in L^1(\Mext)
 \;.
\ee
%
%We require the initial data set to be maximal, $\trg K=0$.

 We will use the Einstein-Maxwell scalar constraint equation, which
for maximal initial data reads
\bel{EMcost} ^{(3)}R = 16 \pi \mu+ |K|_g^2 + 2\Big(|E|_g^2+|B|^2_g
\Big)
 \;,
\ee
where the function $\mu\ge 0$ represents the non-electromagnetic
energy density and $|\cdot|_g$ denotes the norm of a vector with
respect to the metric $g$.

To obtain our inequality we start by bounding \eq{EMcost} from
below, as follows. By~\cite{ChUone} there exists a coordinate
system, with controlled asymptotic behaviour, in which the
metric takes the form
\begin{equation} \label{axmet2}
g = e^{-2U+2\alpha} \left(d\rho^2 + dz^2 \right) + \rho^2 e^{-2U}
\left(d\varphi + \rho\: W_{\rho} d\rho + W_z dz \right)^2 \, .
\end{equation}
Consider an orthonormal frame $e_i$ such that $e_3$ is
proportional to the rotational Killing vector field
$$
 \eta:=\partial_\varphi
 \;.
$$
Let $\theta^i$ denote the dual co-frame; for definiteness we
take
$$
\theta^1 =e^{-U+\alpha} d\rho\;, \quad \theta^2 = e^{-U+\alpha}
dz\;, \quad \theta^3 = \rho e^{-U} \left(d\varphi + \rho W_{\rho}
d\rho + W_z dz \right) \; .
$$
We assume that the initial data are invariant under the flow of
$\eta$; this implies the space-time equations $\mcL_\eta F=0$
and $\mcL_\eta * F=0$,  where a star stands for the Hodge dual,
and $\mcL$ denotes a Lie-derivative. A standard calculation,
together with the hypothesis of simple-connectedness of
$\MUone$, implies then the existence of functions $\chi$ and
$\psi$ such that
\bel{Ernst}
 \partial_\alpha \chi=F_{\mu\alpha} \eta^\mu \;,\qquad
 \partial_\alpha \psi=*F_{\mu\alpha} \eta^\mu  \;.
\ee
We then have
\bean%l{EBlb}
   %\sqrt{\det g}\,\Big(
   |E|_g^2+|B|^2_g
   %\Big)
 & = &
  \frac {e^{2U}}{\rho^2}\Big((\partial_n \chi)^2 +|D\chi|_g^2 +
  (\partial_n \psi)^2 +|D\psi|_g^2 \Big)
  \\
  & \ge &
  \frac {e^{2U}}{\rho^2}\Big(|D\chi|_g^2 +
   |D\psi|_g^2 \Big)
  \nonumber
  \\
  & = &
  \frac {e^{4U-2\alpha}}{\rho^2}\Big((\partial_\rho\chi)^2 +(\partial_z\chi)^2 +
   (\partial_\rho\psi)^2+
   (\partial_z\psi)^2 \Big)
  \;,%
%  \nonumber
%  \\
% &&
\eeal{EBlb}
where $\partial_n$ denotes the derivative in the direction normal to
the initial data hypersurface.

Writing
$$
4\pi T_{\mu\nu}\eta^{\mu}
=F_{\mu\alpha}F_{\nu}{}^{\alpha}\eta^{\mu}-\frac{1}{4}F_{\alpha\beta}F^{\alpha\beta}\eta_\nu
= \partial_{\alpha}\psi F_{\nu}{}^{\alpha}-\frac{1}{4}
F_{\alpha\beta}F^{\alpha\beta}\eta_\nu
 \;,
$$
we are now able to justify
Remark~\ref{J=0}, under the (reasonable) condition that
\bel{29VI.1}
 F_{\mu\nu}\psi = o(r^{-2})
 \;.
\ee
First, the
vanishing of the electric and magnetic charges is immediate. Next,
the symmetry of the problem implies that $\vec J$ is aligned along
the axis of rotation. Letting $J_z$ denote the component of
angular-momentum along the rotation axis, by the Komar identity we
obtain
\begin{eqnarray}
 \nonumber
 16\pi J_z & = &
 \int_{S_\infty}\nabla^{\mu}\eta^{\nu}dS_{\mu\nu}=\frac{1}{2}\int_M \nabla_{\mu}\nabla^{\mu}\eta^{\nu}dS_{\nu}
\\
 \nonumber
& = &\frac{1}{2} \int_M R_{\mu}{}^{\nu}\eta^{\mu}dS_{\nu}=4\pi\int_M T_{\mu}{}^{\nu}\eta^{\mu}dS_{\nu}
\\
& = &
 \nonumber
-\int_{S_{\infty}}
\psi\underbrace{\nabla_{\alpha}F^{\nu\alpha}}_0dS_{\nu}+\int_M
\nabla_{\alpha}(\psi F^{\nu\alpha})dS_{\nu}-\frac{1}{4}\int_{M}
F_{\alpha\beta}F^{\alpha\beta}\eta^\nu dS_{\nu}
\\
& = &   2\int_{S_{\infty}} \psi F^{\nu\alpha}dS_{\nu\alpha}
-\frac{1}{4}\int_{M}F_{\alpha\beta}F^{\alpha\beta}\underbrace{\eta^{\mu}n_{\mu}}_0d^3\mu_g=0
 \;,
 \label{Jvanish}
\end{eqnarray}
where we have used the fact that $\eta$ is tangent to $M$, and
where the first integral in the last line vanishes by
\eq{29VI.1}.

As discussed in~\cite{Dain:2006}, in vacuum the one-form%
\footnote{We take this opportunity to point out a factor of 2
missing in the left-hand-side of Eq.~(2.6) in~\cite{CLW}, which
affects numerical factors in some subsequent equations, but has
no other consequences.}
 \bean
 \label{lambda}
 \twist  &:=& 2
 \epsilon_{ijk}K^j{}_\ell \eta^k \eta^\ell dx^i
 \\
 \nonumber
 &=& 2
  \epsilon({\partial_A,\partial_B,\partial_\varphi})K(dx^B,\partial _\varphi) dx^A
 \\
 &=&
 2  g(\eta,\eta)\epsilon({e_a,e_b,e_3})K(\theta^b,e_3) \theta^a
% \;.
% \\
% &= &
% g(\eta,\eta)\epsilon_{A}{}^{B}K(\eta,e_B)\theta^A=
\eeal{twist}
is closed. Here, as before, the upper case indices $A,B=1,2$
correspond to the coordinates $(\rho, z)$, while the lower case
indices $a,b=1,2$ are frame indices. In electro-vacuum we have
instead (see, e.g.,~\cite{Weinstein3})
\bel{twist2}
  d\Big(\twist-2(\chi d\psi - \psi d\chi)\Big)=0
 \;,
\ee
%$$
%
and, since we have assumed that $\MUone$ is simply connected, there
exists a function $\myomega$ such that
\bel{twistex}
  \twist  = 2 (d\myomega + \chi d\psi - \psi d\chi)
 \;.
\ee
Then, writing $K_{b3}$ for $K(e_b,e_3)$, and using
$\epsilon_{ab}:=\epsilon(e_a,e_b,e_3)\in \{0,\pm 1\}$, we have
$$
2\rho^2 e^{-2U}(K_{23}\theta^1- K_{13}\theta^2)= \twist_\rho  d\rho +
\twist_z  dz
 \;;
$$
equivalently
\bel{Komeq} K_{13}=-\frac{e^{3U-\alpha}}{2\rho^2}\twist_z \;,
 \quad
 K_{23}=\frac{e^{3U-\alpha}}{2\rho^2}\twist_\rho \;,
\ee
so that
%, since $\sqrt{\det g}= \rho e^{-3U+2\alpha}$,
% We then have
%
\bel{Klb}
 %\sqrt{\det g}
e^{2(\alpha-U)}\,|K|_g^2 \ge 2
 e^{2(\alpha-U)}
 %\sqrt{\det g}
(K_{13}^2+K_{23}^2) = \frac {e^{4U}} {2\rho^4} |\twist
|^2_\delta
 \;.
\ee

In~\cite{ChUone}
(compare~\cite{Brill59,GibbonsHolzegel,Dain:2006}) it has been
shown that
\bean m&=& \frac{1}{16 \pi} \int \Big[\phantom{}^{(3)}R +
\frac{1}{2} \rho^2
  e^{-4\alpha+2U}\left(\rho W_{\rho, z} - W_{z,\rho}\right)^2 \Big] e^{2(\alpha -U)} d^3 x
 \\
 & & +
\frac{1}{8\pi}\int \left(D U\right)^2 d^3x \, .
 \label{mf}
\eea
Inserting \eq{EBlb} and \eq{Klb} into \eq{mf} we
obtain
\begin{eqnarray}
 \nonumber m& \ge & \frac{1}{16 \pi} \int
\Big[\phantom{}^{\, (3)}R  e^{2(\alpha-U)}+   2\left(D U\right)^2
\Big]d^3x
 \\
 & \ge &
\frac{1}{8 \pi} \int \Big[ \left(D U\right)^2 +\frac {e^{4U}}
{\rho^4} \left(D \myomega+\chi D\psi - \psi D \chi \right)^2 +\frac
{e^{2U}} {\rho^2} \left((D  \chi)^2+ (D\psi)^2\right)\Big]d^3x
 \, ,
 \nonumber
 \\
 &&
  \label{mf2}
\end{eqnarray}
where, from now on, we use the symbol $Df$ to denote the
gradient of a function $f$ with respect to the flat metric
$\delta$, and where $(Df)^2\equiv |Df|^2_{\delta}$.

It follows from \eq{Ernst} that $\psi$ and $\chi$ are constant
on each connected component $\mcA_j$ of the ``axis"
$$
 \mcA:=\{\rho=0\}\setminus\{z=0\}\;;
$$
\eq{twistex}-\eq{Komeq} then
show that so is $\myomega$. We set
\bel{AxisValue}
 \myomega_j:= \myomega|_{\mcA_j}
 \;, \quad \psi_j:=\psi|_{\mcA_j}
 \;, \quad \chi_j:=\chi|_{\mcA_j}
 \;.
\ee

We have the following, from which Theorem~\ref{TM1} immediately
follows:

\begin{Theorem}
 \label{TM2}
Let $(M,g,K,   v,   \chi,\psi)$ be a three dimensional smooth
data set invariant under an action of $\Uone$, where $M$ is the
union of a compact set and of two asymptotically flat regions
$M_1$ and $M_2$, in the sense of \eq{falloff1} together
with~\eq{KFfalloff1}, and where $v,\psi$ and $\chi$ are global
potentials as in \eq{twistex}-\eq{Komeq} and \eq{Ernst}. Assume
that \eq{EMcost} holds with $\mu\ge 0$.  Let $m$ and $\vec J$
denote the ADM mass and angular momentum of $M_1$, and let
$Q_E$ and $Q_B$ be the global electric and magnetic charges of
$M_1$. If $\MUone$ is simply connected, then
$$
 m \ge \sqrt{\frac{|\vec J|^2}{m^2}+Q_E^2+Q_B^2} \;.
$$
\end{Theorem}
The proof of Theorem~\ref{TM2} proceeds as follows. If the mass
is infinite there is nothing to prove; under the current
hypotheses this will be the case iff $^{(3)}R\not \in L^1(M)$.
Otherwise, in view of \eq{mf2}, one considers the action
\begin{equation} I:=  \int \Big[ \left(D U\right)^2 +\frac {e^{4U}}
{\rho^4} \left(D \myomega+\chi D\psi - \psi D \chi \right)^2 +\frac
{e^{2U}} {\rho^2} \left((D  \chi)^2+ (D\psi)^2\right)\Big]d^3x
  \label{action}
 \;.
\end{equation}
One wishes to show that $I$ is bounded from below by  the
right-hand-side of our bound, which is the value of the
corresponding action for the extreme Kerr-Newman solution with
the same global Poincar\'e and Maxwell charges. The proof can
be obtained by following the argument in~\cite{CLW}, except for
supplementary difficulties in several places related to the new
structure of the term $\frac {e^{4U}} {\rho^4} \left(D
\myomega+\chi D\psi - \psi D \chi \right)^2$. For example, when
a cut-off function $\varphi_\eta$ is used in~\cite{CLW}, the
estimates do not appear to go through using the arguments
there. This can be circumvented as follows: Let the quantities
decorated with tildes refer to the extreme Kerr-Newman
solution. Let $\theta=U,v,\chi,\psi$ and write
$$\theta_{\jmu }:=\varphihere_{\jmu }\theta+(1-\varphihere_{\jmu })\,\tilde\theta=\varphihere_{\jmu }(\theta-\tilde\theta)+\tilde\theta\;.$$
Set
\bel{lambdasig}\lambda_{\,\jmu } :=  D v_{\jmu }+\chi_{\jmu } D
\psi_{\jmu }-\psi_{\jmu } D\chi_{\jmu }\;. \ee
While a direct estimate of $\lambda_{\jmu}$  in the most
singular integrals in~\cite{CLW} does not appear to be
immediate, it turns out that each summand in the identity
\begin{eqnarray}
 \nonumber
\lambda_{\jmu }
&=& \varphihere_{\jmu }\lambda + (1-\varphihere_{\jmu })\tilde\lambda
 +  D \varphihere_{\jmu }(v-\tilde v)+ D \varphihere_{\jmu }(\tilde\chi\psi-\tilde\psi\chi)
\\
\label{goodex} && + \varphihere_{\jmu }(1-\varphihere_{\jmu
})\left\{(\psi-\tilde\psi) D(\chi-\tilde\chi)-(\chi-\tilde\chi)
D(\psi-\tilde\psi)\right\}
%\;.
\end{eqnarray}
%$\lambda_\jmu$
can be handled in a way similar to the original integrals
in~\cite{CLW}.  For example, one of the steps of the proof is
to establish  that
\bel{intto0}
\int_{\eta/2<r<\eta}\frac{e^{4U_{\eta}}}{\rho^4}|\lambda_{\eta}|^2d^3x\rightarrow_{\eta\rightarrow
0} 0\;, \ee
where $r^2=\sqrt{\rho^2+z^2}$. Recall that near $r=0$  the
coordinates $(\rho,z)$ can be obtained from the usual
cylindrical coordinates in the other asymptotically flat
region, which we denote by $(\hat \rho, \hat z)$, by an
inversion $(\hat\rho,\hat z)=(\frac{\rho}{r^2},\frac{z}{r^2})$,
compare~\cite[Theorem~2.9, p.~2580]{ChUone}. This leads to
estimates for small $r$, equivalently for large $\hat r$, such
as
\bel{chi origin} |D\chi|_{\delta}= \frac{1}{r^2}|\hat D\chi|_{\hat
\delta} \lesssim \frac{1}{r^2}\hat\rho\hat r^{-\gamma-1}=\rho
r^{\gamma-3}\;,\,\,r\rightarrow 0\;. \ee
From this and the known asymptotic behaviour of extreme
Kerr-Newman one obtains, when the decay exponent $\lambda$ of
$K$ (see \eq{falloff1}) satisfies $\lambda\ge
2\gamma+1$,%
\footnote{For $\lambda\leq 2\gamma+1$ the dominating behaviour
in~\eqref{twistex} is governed by $\lambda$, which leads to
$v-\tilde v=O(r^{\lambda-3})$ and the necessity to impose
$\lambda>5/2$, as in vacuum~\cite[p.~2602]{CLW}.}
\bel{v-v origin} v-\tilde v=O(r^{2\gamma-2})\;.\ee
This allows us to estimate the contribution of the term $ D
\varphihere_{\jmu }(v-\tilde v)$ of~\eqref{lambdasig} in the
subregion $\rho\geq z$ of the integral \eq{intto0} as
$$
 \int_{\eta/2}^{\eta}\frac{r^4}{\rho^4r^2}\left(r^{2\gamma-2}\right)^2r^2dr
 =O(\eta^{4\gamma-3})\to_{\eta\to 0} 0 \ \text{provided that} \ \gamma>3/4
\;.
$$
This explains our ranges of $\lambda$ and $\gamma$
in~\eqref{falloff1} and~\eqref{KFfalloff1}.

A detailed presentation, with some simplifications of the
argument of~\cite{CLW}, will be given
elsewhere~\cite{JLCCharge,CostaPhD}.

\bigskip

{\sc Acknowledgements:} We are grateful to S.~Dain and E.~Delay
for many useful discussions, and for sharing their calculations
of the potentials for the extreme Kerr-Newman solution.

\bibliographystyle{amsplain}
\bibliography{../references/hip_bib,%
../references/reffile,%
../references/newbiblio,%
../references/newbiblio2,%
../references/bibl,%
../references/howard,%
../references/bartnik,%
../references/myGR,%
../references/newbib,%
../references/Energy,%
../references/netbiblio}
\end{document}